\documentclass[a4paper]{jpconf}
\usepackage{graphicx}
\usepackage[centertags]{amsmath}
\usepackage{amsfonts}
\usepackage{amssymb}

\usepackage{amsthm}
\begin{document}
\title{Birefringence in pseudo--Finsler spacetimes}

\author{Jozef Sk\'akala and Matt Visser}

\address{School of Mathematics, Statistics, and Computer Science,\\
Victoria University of Wellington, Wellington, New Zealand}

\ead{\{jozef.skakala,matt.visser\}@mcs.vuw.ac.nz}

\begin{abstract}
Based on the analogue spacetime programme, and many other ideas currently mooted in ``quantum gravity'', there is considerable ongoing speculation that the usual pseudo-Riemannian (Lorentzian) manifolds of general relativity might eventually be modified at short distances.  Two specific modifications that are often advocated are the adoption of Finsler geometries (or more specifically, pseudo-Finsler spacetimes) and the possibility of birefringence (or more generally, multi-refringence). We have investigated the possibility of whether it is possible to usefully and cleanly deal with these two possibilities \emph{simultaneously}. That is, given two (or more) ``signal cones'': Is it possible to naturally and intuitively construct a  ``unified'' pseudo-Finsler spacetime such that the pseudo-Finsler metric is null on these ``signal cones'', \emph{but has no other zeros or singularities}? Our results are much less encouraging than we had originally hoped, and suggest that while pseudo-Finsler spacetimes are certainly useful constructs, it is physically more appropriate to think of physics as taking place in a single topological manifold that carries several distinct pseudo-Finsler metrics, one for each polarization mode.
\end{abstract}

\newcommand{\A}{{\cal A}}
\newcommand{\h}{{\cal H}}
\newcommand{\s}{{\cal S}}
\newcommand{\W}{{\cal W}}
\newcommand{\BH}{\mathbf B(\cal H)}
\newcommand{\KH}{\cal  K(\cal H)}
\newcommand{\Real}{\mathbb R}
\newcommand{\Complex}{\mathbb C}
\newcommand{\Field}{\mathbb F}
\newcommand{\RPlus}{[0,\infty)}
\newcommand{\norm}[1]{\left\Vert#1\right\Vert}
\newcommand{\essnorm}[1]{\norm{#1}_{\text{\rm\normalshape ess}}}
\newcommand{\abs}[1]{\left\vert#1\right\vert}
\newcommand{\set}[1]{\left\{#1\right\}}
\newcommand{\seq}[1]{\left<#1\right>}
\newcommand{\eps}{\varepsilon}
\newcommand{\To}{\longrightarrow}
\newcommand{\RE}{\operatorname{Re}}
\newcommand{\IM}{\operatorname{Im}}
\newcommand{\Poly}{{\cal{P}}(E)}
\newcommand{\EssD}{{\cal{D}}}
\theoremstyle{plain}
\newtheorem{thm}{Theorem}[section]
\newtheorem{cor}[thm]{Corollary}
\newtheorem{lem}[thm]{Lemma}
\newtheorem{prop}[thm]{Proposition}
\theoremstyle{definition}
\newtheorem{defn}{Definition}[section]
\theoremstyle{remark}
\newtheorem{rem}{Remark}[section]
%
\renewcommand{\theequation}{\thesection.\arabic{equation}}
\def\d{{\mathrm{d}}}
\section{Introduction}
Motivation for our work~\cite{Skakala} comes from various speculations that came up mainly from:
\begin{itemize}
\item 
the analogue spacetime programme~\cite{Analogue};
\item
 various different approaches to ``quantum gravity'',
 \\
  (loop quantum gravity~\cite{LQG}, string models~\cite{string}, causal dynamical triangulations~\cite{causal});
\item
various different approaches to ``quantum gravity phenomenology'',
\\ (DSR~\cite{DSR}, VSR~\cite{VSR},  noncommutative geometry~\cite{ncg});
\item   
various different approaches to ``exotic'' cosmologies~\cite{VSL, superluminal}.
\end{itemize}
Many of these approaches strongly suggest the possibility of high energy Lorentz symmetry breaking~\cite{Jacobson,Lorentz} and/or birefringence (or multi-refringence)~\cite{birefringence}.

If we wish to create a geometric theory encoding causal relations  more complicated  than those of special relativity, then we must unfortunately depart from the standard concept of pseudo-Riemannian geometry, and adopt something more general. This more general geometry is almost inevitably a non-positive definite analogue of Finsler geometry~\cite{Finsler, Antonelli, Duval}, that is,  a geometry which can most descriptively  be called a pseudo-Finsler geometry~\cite{pseudo-Finsler}. (One might also try to generalize the spacetime in a slightly different direction by adopting the notion of an ``area metric''~\cite{area}, but we shall not take that particular route for now.) 

One particularly attractive route to constructing Lorentz symmetry breaking geometries is to consider analogue spacetimes. These are classical systems providing us,  at the macroscopic level, with emergent phenomena similar to gravity, while being radically different at the microscopic level --- and at the same time are physically well understood~\cite{Analogue,Lorentz}. They provide us with analogues of many aspects of quantum gravity phenomenology, and are also possible sources of experimental tests for quantum gravity phenomenology. (We can even hope for the detection of Hawking radiation~\cite{Analogue}.)

In our specific case~\cite{Skakala} we chose a particular analogue spacetime model --- that of a birefringent crystal~\cite{Born} --- providing us with explicit Lorentz symmetry breaking relations, which at the same time has a simple enough structure (the dispersion relations factorize, in general, to give a two-cone structure) so as to give us a chance to create in an intuitive way a full spacetime pseudo-Finsler geometry (with both metric and co-metric structures). 

Birefringent crystals have in the past traditionally been analysed in terms of 3-dimensional \emph{spatial} Finsler geometry. (See, for instance,  Antonelli, Ingarden, and Matsumoto~\cite{Antonelli}, and Duval \cite{Duval}.) But  these analyses are purely \emph{spatial} --- they effectively pick a fixed inertial reference frame, the rest frame of the crystal, and hence pick a preferred $3d$ spacelike slices, so these approaches do not cross the boundaries of classical nonrelativistic mechanics. In contrast, we wish to construct a full $4d$ \emph{spacetime} pseudo-geometry based on birefringent crystals --- a fully $(3+1)$-dimensional construction.

\section{Basics of pseudo-Finsler geometry}

How should the theory of crystal optics provide us with a Lorentz symmetry breaking geometry? The theory gives us a clear ideas how the waves propagate, specified by the possible four-wavevectors $k = (\omega; \mathbf{k})$ with their corresponding phase velocities $\mathbf{v}_p=\omega \,\mathbf{k}/||\mathbf{k}||^2 = \omega \, \hat{\mathbf{k}}/||\mathbf{k}||$,  and a clear idea about how the energy is transferred, specified by the possible group three-velocities $\mathbf{v}_g$. The group three-velocity can most naively be extended to a group four-velocity by the prescription $v=(1; \mathbf{v}_g)$. More generally, if we choose to parameterize the integral curves of the group velocity by some arbitrary parameter $s$, then the possible group four-velocities become  $v=(1; \,\mathbf{v}_g) \times [\d t/\d s] =  (v^0; \,\mathbf{v}) \propto (1; \,\mathbf{v}_g)$.  These objects can be explicitly interpreted as describing how quanta of electromagnetic energy (hence photons) propagate. Note that any multiple of a possible  four-wavevector is again a possible four-wavevector, and similarly by construction any multiple of a possible group four-velocity is again a possible group four-velocity. 

So the spacetime geometry we are seeking should provide us with a pair of pseudo-norms $N(v)$ and $N'(k)$ specifying as null four-vectors exactly the possible group four-velocities and four-wavevectors respectively. That is, the pseudo-norm should give us the relation $N(v)=0$ or $N'(k)=0$ for the possible group four-velocities/four-wavevectors, and \emph{only} for the possible group four-velocities/four-wavevectors. 

Let us now precisely state the conditions we should impose on a pseudo-norm in order for it to be physically interesting: A pseudo-norm $N(v)$ on four-vectors is a map from $V\to i\mathbb{R_{+}}\cup\mathbb{R_{+}}\cup{0}$, which is~\cite{Skakala}:
\begin{itemize}
\item[a)]  1-homogeneous,
\item[b)] the square $N^{2}(v)$ is smooth on $V/\{0\}$, and 
\item[c)] the gradient
$\partial_{v_{i}}[N^{2}(v)]$ is a bijection from $V\to V^*$,   the vector-space dual to $V$.
\end{itemize}
This concept of pseudo-norm now specifies what we mean by pseudo-Finsler geometry~\cite{Skakala}.

To see that we can build up a full spacetime geometry out of this concept, we shall define a few other geometric concepts in complete analogy to the traditional positive definite Finsler case.
The metric tensor is defined, as usual,  by
\begin{equation}
g_{i j}(v)=\frac{1}{2}\;\frac{\partial^{2}[N^{2}(v)]}{\partial v^{i}\; \partial v^{j}}.
\end{equation}
We see that this definition reduces to standard pseudo-Riemann case for
\begin{equation}
N(v)=\sqrt{g_{i j}\;v^{i}\; v^{j}}
\end{equation}
where (in this case) $g_{i j}$ does not depend on $v$. By using Euler's theorem, which states that for $n$-homogeneous functions
\begin{equation}
\frac{\partial f(v)}{\partial v^{i}}\; v^{i}=n\,f(v),
\end{equation}
we can easily define a mapping from vectors (four-velocities) to co-vectors (four-wavevectors) as 
\begin{equation}
k_i(v)=\frac{\partial [N^{2}(v)]}{\partial v^{i}} = g_{i j}(v)\;v^{j}.
\end{equation} 
Then by using a Legendre transform, we
can easily define the co-Finsler pseudo-norm $N'(k)$ as
\begin{equation}
N'(k(v))=N(v)
\end{equation}
and co-Finsler metric tensor as 
\begin{equation}
g^{i j}(k(v))=\frac{1}{2}\;\frac{\partial^{2} [N'^{2}(k(v))]}{\partial k^{i}(v) \; \partial k^{j}(v)}.
\end{equation}
Then it can be proven that 
\begin{equation}
v^i(k)=\frac{\partial [N'^{2}(k)]}{\partial k_{i}} = g^{ij}(k) \; k_j,
\end{equation}
where $v(k)$ and $k(v)$ are inverse mappings: $v(k(v))=v$. 
This shows that we have a meaningful generalization of pseudo-Riemann geometry.
Note that we can easily define a pseudo-Finsler connection, and hence curvature, in this formalism simply by extremalizing the Finsler distance ($v = \d x/\d s$):
\begin{equation} 
\ell(a,b) = \int_{a}^{b}\sqrt{g_{i j}(v)\; v^{i}\; v^{j}}\; \d s.
\end{equation}

\section{Birefringent crystal}
\subsection{General constraints}
What are the specific physical constraints given by birefringent crystals on our geometry? We consider it meaningful to impose
the following constraints:
\begin{itemize}
\item[a)] The pseudo-norm should, for vectors approaching the purely timelike $(a,\mathbf{0})$ vector (in the fixed coordinates ``at rest'' with respect to the crystal), approach the pseudo-Riemann flat-space pseudo-norm.
 \item[b)] The pseudo-norm should break Lorentz invariance as in a birefringent crystal.
 \item[c)] Each ``constant proper time'' hypersurface, specified by $N^{2}(v)=-\tau_0^{2}$, should be connected, and should contain the point $(a,\mathbf{0})$.
 \end{itemize}
Let us discuss these constraints a little bit: The first is just requirement that we want to recover low-energy physics as we know it (in a pseudo-Riemann form).
The second is trivial: it is the specific contribution of using birefringent crystal as an analogue model (unlike the first and the third condition, which can be considered as generic).
The third condition states that if we have a massive particle, it should not be able to have two distinct sets of four-velocities between which it is not able to undergo a smooth transition. It also means that any massive particle should be able to move arbitrarily slowly in the rest frame of the crystal.

\subsection{Classical theory of birefringent crystals giving rise to co-metric structure}
Now focus on the classical theory of birefringent crystals (see \cite{Born} for a detailed exposition). It is typically formulated in a preferred coordinate frame, which
is the inertial frame of the crystal,  with spatial coordinate axes aligned with the principal axes of the crystal. In fact in the rest fame the crystal is described by a $3\times3$ dielectric tensor $\epsilon_{i j}$, where the principal axes are defined as axes where the dielectric tensor is diagonalized.

The relation between the components of the possible four-wavevectors can be extracted from the Fresnel equation for wave normals~\cite{Skakala, Born}:
\begin{equation}
 1=\sum_{i=1}^{3}\frac{k^{2}_{i}}{\mathbf{k}^{2}-\omega^{2}\mu\epsilon_{i}}.
 \end{equation}
This factorizes  into
\begin{equation}
F'(k) =F'_{1} \; F'_{2}=
\left(\omega^{2}+b(\mathbf{k})-\sqrt{b^{2}(\mathbf{k})-c(\mathbf{k})}\right) \; 
\left(\omega^{2}+b(\mathbf{k})+\sqrt{b^{2}(\mathbf{k})-c(\mathbf{k})}\right)=0,
\end{equation}
where
\begin{equation}
b(\mathbf{k})=-\frac{
k_{1}^{2}\,\epsilon_{1}(\epsilon_{2}+\epsilon_{3})+k_{2}^{2}\,\epsilon_{2}(\epsilon_{1}+\epsilon_{3})+k_{3}^{2}\,\epsilon_{3}(\epsilon_{1}+\epsilon_{2})
}{2c^{2}\mu\epsilon_{1}\epsilon_{2}\epsilon_{3}},
\end{equation}
and
\begin{equation}
 c(\mathbf{k})=\frac{\sum_{i=1}^{3}k_{i}^{4}\,\epsilon_{i}
 +k_{1}^{2}k_{2}^{2}\,(\epsilon_{1}+\epsilon_{2})+k_{1}^{2}k_{3}^{2}\,(\epsilon_{1}+\epsilon_{3})+k_{2}^{2}k_{3}^{2}\,(\epsilon_{2}+\epsilon_{3})
 }{c^{4}\mu^{2}\epsilon_{1}\epsilon_{2}\epsilon_{3}}.
 \end{equation}
As we see,  in the general case this relation encodes a two-cone structure, one for each polarization, each taking in general a distinct
phase velocity $\mathbf{v}_p=\omega\,\mathbf{k}/||\mathbf{k}||^2 = \omega\,\hat{\mathbf{k}}/||\mathbf{k}||$. So we see that here the space-time causal structure is much more complicated than in special relativity.

\subsection{Co-metric structure construction}
Since the four-wavevector is in fact a covector, these relations give rise to a co-Finsler structure rather then to a Finsler structure.
Hence they give us a generic candidate for the co-pseudo-Finsler norm, which can be written as $N'(k)^2\propto F'(k)$ or more precisely  $N'(k)^2=M'(k)\,F'(k)$, where $M'(k)$ is an otherwise arbitrary function fulfilling following constraints:
\begin{itemize}
\item[a)] 
$\displaystyle \frac{\partial[M'(k)F'(k)]}{\partial k^{i}}$ should be a bijection from $V^*\to V$;
\item[b)]
$M'(k)$ is a $-2$ homogeneous map to $\mathbb{R}$;
\item[c)]
 smooth on $V^*/\{0\}$;
\item[d)]
 inside both signal cones negative, outside both positive, between them nonzero;
\item[e)]
$M'(k)\,F'(k)$ should approximate the  flat-space pseudo-Riemann norm (squared) in the given coordinates for $k$ close to $(a,\mathbf{0})$ and such
vectors should lie on some hypersurface (mass-shell) given by: $M'(k)\, F'(k)=-m^{2}$, which should be always connected.
\end{itemize}
These conditions follow trivially from the conditions we imposed on any physically meaningful Finslerian geometry.

\subsection{Classical theory of birefringent crystals giving rise to metric structure}
We should also consider the Finsler pseudo-norm; this should be extracted from the equations describing how energy propagates, which is the equation
determining how photons propagate (their group velocities).  This should be put into the form $g_{\mu \nu}(v) \; \d x^{\mu}\; \d x^{\nu}=0$, so that  it encodes the pseudo-Finsler metric tensor as opposed to co-Finsler metric tensor. The information regarding energy propagation is given by the Fresnel ray equation,
which is:
\begin{equation}
\frac{t^{2}_{1}}{\frac{1}{v^{2}_{r}}-\frac{1}{v^{2}_{2}}} + \frac{t^{2}_{3}}{\frac{1}{v^{2}_{r}}-\frac{1}{v^{2}_{1}}}+\frac{t^{2}_{3}}{\frac{1}{v^{2}_{r}}-\frac{1}{v^{2}_{3}}}=0.
\end{equation}
Here
\begin{equation}
v^{2}_{i}=\frac{1}{\mu\epsilon_{i}}.
\end{equation}
Using the four vector $x=(t; \mathbf{x})$, the Fresnel ray equation can be transformed into
\begin{equation}
F(x)=F_{1}(x)\; F_{2}(x)=
\left(t^2+b(\mathbf{x})-\sqrt{b^{2}(\mathbf{x})-c(\mathbf{x})}\right)\;
\left(t^2+b(\mathbf{x})+\sqrt{b^{2}(\mathbf{x})-c(\mathbf{x})}\right)=0,
\end{equation}
where
\begin{equation}
b(\mathbf{x})=
-c^{2}\frac{v^{2}_{1}(v^{2}_{z}-v^{2}_{2})\,x^{2}_{1}+v^{2}_{2}(v^{2}_{1}+v^{2}_{3})\,x^{2}_{2}+v^{2}_{3}(v^{2}_{1}+v^{2}_{2})\,x^{2}_{3}}
{2v^{2}_{1}v^{2}_{2}v^{2}_{3}},
\end{equation}
and
\begin{equation}
c(\mathbf{x})=\frac{c^{4}\mathbf{x}^{2}\; (\sum^{3}_{i=1}v^{2}_{i}x\,^{2}_{i})}{v^{2}_{1}v^{2}_{2}v^{2}_{3}}.
\end{equation}
Obviously the  the ray equation also encodes a two-cone structure. 

\subsection{Metric structure construction and interconnecting the two structures}
The generic Finsler pseudo-norm is in this case expressed exactly in the same way as was the  co-Finsler pseudo-norm in the previous case,
by the function $N(v)^2=M(v)\; F(v)$, where $M(v)$ is again an arbitrary function fulfilling exactly the same conditions as the $M'(k)$ function, we just have to exchange $V^*$ for $V$.

Now the last step in putting constraints on geometric construction is to relate the Finsler and co-Finsler structures. The items which are as yet 
undetermined  are the functions $M(v)$, and $M'(k)$. So they have to fulfill the last condition, which is the condition of forming
the full united geometry. This condition is given by the Legendre transform relation:
\begin{equation}
\left[N'\left(\frac{\partial [N(v)^2]}{\partial v^{i}}\right)\right]^2=N(v)^2.
\end{equation}
In our language, the function $M(v)$ must be connected to $M'(k)$ by:
\begin{equation}
M'\left(\frac{\partial [M(v)F(v)]}{\partial v^{i}}\right) \;
F'\left(\frac{\partial [M(v)F(v)]}{\partial v^{i}}\right)=M(v) \; F(v).
\label{E:compatibility}
\end{equation}
Thus we see that if we have somehow  found an  appropriate $M(v)$, then $M'(k)$ will be uniquely determined.

To find solutions of such an abstract (although precisely formulated) mathematical exercise is not an easy task. At this early stage it is quite doubtful whether general solutions exist. We shall look at particular cases, and try to check some intuitive solutions, since our original objective was to build up geometry in an intuitive way.

\section{Attempts to solve our problem and conclusions}
The simplest case of crystal optics is an isotropic medium $v_{1}=v_{2}=v_{3}=v_o$. The obvious interpretation in this case is as a pseudo-Riemann
metric, but with $v_o$ playing the role of the speed of light~\cite{Skakala,Born}. (Strictly speaking, even this obvious solution does not fulfill our condition (e), unless $v_o=c$, which implies there is no crystal at all.)

In a uniaxial crystal ($v_{1}=v_{2}\neq v_{3}$),  the $F(v)$ and $F'(k)$ functions decompose
into the product of two Riemannian structures~\cite{Skakala}, but simple and intuitive ansatz\"e for the $N(k)$ and $N'(k)$ functions lead to serious difficulties.
For instance a suggestion where either
\begin{equation}
N(v)^2= \sqrt{F(v)} = \sqrt{F_{1}(v)\cdot F_{2}(v)} \qquad \hbox{or} \qquad N'(k)^2=\sqrt{F'(k)}=\sqrt{F'_{1}(k)\cdot F'_{2}(k)}
\end{equation}
(with the two relations being related by Legendre transformation)
encounters problems with condition (e), the putative Finsler norms are nonreal between the two signal cones,
and not smooth at the signal cones. (Even second derivatives do not exist there, hence the pseudo-(co)-Finsler metric tensor is undefined there~\cite{Skakala}).

In the case of biaxial crystal ($v_{1}\neq v_{2}\neq v_{3}\neq v_{1}$)  there are even bigger problems with this ``solution" --- even the individual $F_{1}(v)$, $F_{2}(v)$ and $F'_{1}(k)$, $F'_{2}(k)$ are not themselves pseudo-Riemannian structures; they are nonsmooth at the optical axes~\cite{Skakala}.  Obviously, the previous problems with the ``unified'' $\sqrt{F_{1}\cdot F_{2}}$ or Legendre-equivalent  $\sqrt{F'_{1}\cdot F'_{2}}$ structures remain.

We could try to work out more sophisticated solutions to our ``compatibility problem'' (\ref{E:compatibility}), and at least prove existence. We have found a partial solution for this problem, except for proving the bijection condition, which is quite tricky. But the fact that this construction lacks any intuitive basis is quite disturbing. In these circumstances we have to accept the unwelcome conclusion that although birefringent crystal have given us many ideas about the construction of pseudo-Finsler geometries, they have not provided
us with a clear and intuitive example of  a ``unified'' geometry capable of handling both polarization modes simultaneously. So the original aim we posed at the beginning of this article has not been accomplished. 

In conclusion: Our results are much less encouraging than we had originally hoped, and suggest that while pseudo-Finsler spacetimes are certainly useful constructs, in birefringent situations it does not appear possible to naturally and intuitively construct a  ``unified'' pseudo-Finsler spacetime such that the pseudo-Finsler metric is null on both ``signal cones'', but has no other zeros or singularities --- it seems physically more appropriate to think of physics as taking place in a single topological manifold that carries distinct pseudo-Finsler metrics, one for each polarization mode.

\section*{References}

\end{document}